\documentclass[twocolumn,floatfix,superscriptaddress,aps,prb]{revtex4-1}
\usepackage{graphicx}
\usepackage{amsmath}
\usepackage{amssymb}
\usepackage{amsfonts}
\usepackage{amstext}

\usepackage{float}
\usepackage[english]{babel}

\usepackage{bm}
\usepackage{hyperref}
\usepackage{braket}
\usepackage{url}
\usepackage{tcolorbox}
\usepackage{physics}
\usepackage{longtable}
\usepackage{mathtools}
\usepackage{tensor}
\usepackage{import}
\usepackage[misc]{ifsym}
\usepackage{dsfont}
\usepackage{color}
\usepackage{ulem}

\usepackage{xcolor}
\hypersetup{
    colorlinks,
    linkcolor={red!50!black},
    citecolor={blue!50!black},
    urlcolor={blue!80!black}
}

\numberwithin{equation}{section}

\newcommand{\cd}{\hat{c}^{\dagger}}

\newcommand{\ad}{\hat{a}^{\dagger}}
\newcommand{\bd}{\hat{b}^{\dagger}}
\newcommand{\cn}{\hat{c}}

\newcommand{\an}{\hat{a}}
\newcommand{\bn}{\hat{b}}

\newcommand{\e}{\varepsilon}
\newcommand{\lb}{\left(}
\newcommand{\rb}{\right)}
\newcommand{\adup}{\hat{a}^{\uparrow\dagger}}
\newcommand{\addo}{\hat{a}^{\downarrow\dagger}}
\newcommand{\anup}{\hat{a}^{\uparrow}}
\newcommand{\ando}{\hat{a}^{\downarrow}}
\newcommand{\bdup}{\hat{b}^{\uparrow\dagger}}

\newcommand{\bnup}{\hat{b}^{\uparrow}}

\newcommand{\hup}{h^{\uparrow}}

\newcommand{\pp}{p'}
\newcommand{\qp}{q'}

\begin{document}

\title{Entanglement correction due to local interactions in many-body systems}

\author{Yevheniia Cheipesh}
\affiliation{Instituut-Lorentz, Leiden, The Netherlands, email: \href{mailto:gene.cheypesh@gmail.com}{gene.cheypesh@gmail.com}}
\author{Lorenzo Cevolani}
\affiliation{Institut f\"ur Theoretische Physik,
 Georg-August-Universit\"at G\"ottingen, Germany}
\affiliation{
 Graphcore, Oslo, Norway
}
\author{Stefan Kehrein}
\affiliation{Institut f\"ur Theoretische Physik,
 Georg-August-Universit\"at G\"ottingen, Germany}

\begin{abstract}
The correction to the area law for the bipartite min-entanglement entropy of weakly and locally interacting fermions is calculated based on a perturbative extension of the flow equation holography method \cite{kehrein2017flow}. Explicit calculations for the one- and two-dimensional case (and similarly for higher dimensions) show that the leading correction to the entanglement entropy of non-interacting fermions up to  $U^2$ in the interaction strength does not change the scaling, but only affects the pre-factor of the leading logarithmic term multiplying it by the quasiparticle residue. A term sub-leading to the area law is also present. It is proportional to $U^2$ in the interaction strength and scales linearly with the system size. 
\end{abstract}


\maketitle
\textit{Introduction---}
One of the most general and informative measures of quantum correlations is the entanglement entropy. \cite{horodecki2009quantum, henderson2001classical} In quantum mechanics, positive entropy may arise even without an objective lack of information, which is the source for classical entropy. \cite{shannon1948note, article}

Entanglement entropy characterises the macroscopic phases of matter, its criticality,\cite{vidal2003entanglement,calabrese2004entanglement, barthel2006entanglement} many-body localisation phenomena\cite{bardarson2012unbounded} and even the topological phase\cite{kitaev2006topological, jiang2012identifying, fidkowski2010entanglement} in a very abstract manner, without knowing a lot about the microscopic details of the system. The famous Ryu-Tagayanagi conjecture \cite{ryu2006aspects} also offers a way to better understand AdS-CFT duality as an explicit realization of the holographic principle.\cite{kehrein2017flow, kabat1995black,solodukhin2011entanglement,ryu2006aspects}. It posits a quantitative relationship between the entanglement entropy of conformal field theory and the geometry of an associated Anti-de Sitter spacetime. In such settings, one usually encounters a so-called area law for the scaling of the entanglement entropy with the system size. Such a scaling often appears in the ground state of locally interacting systems, which is the most typical kind of interaction.\cite{eisert2008area}.

Entanglement entropy can also serve as an identification of quantum chaotical systems.\cite{liu2018quantum, bandyopadhyay2002testing, wang2004entanglement}. Another interesting case of the scaling is a so-called volume law when the entanglement entropy saturates. This case is interesting because it is inherent to quantum chaotical systems \textit{e.g.} Sachdev-Ye-Kitaev model, as well and thermal and excited states.\cite{eisert2008area, shiba2014volume, serbyn2013universal}

The entanglement behaviour for systems in the ground state has been studied extensively\cite{eisert2008area, horodecki2009quantum, calabrese2004entanglement,cramer2006entanglement, cheipesh2019exact}, whereas, despite the various studies~\cite{wang2014renyi,pari2019quasiparticle, serbyn2013universal,gong2017entanglement,barthel2006entanglement,berkovits2012entanglement}excited states or  even the ground state of the interacting systems pose many questions that still go unanswered \cite{castro2018entanglement}. Unfortunately, not only is the exact analytical study usually an impossible task, but numerical simulations can also be very challenging, especially for exotic systems in the critical phases.

In this paper, we investigate the theoretical tool that allows  to analytically study the min-entanglement entropy not only of a free many-body system but also an interacting one without any restrictions on the dimensionality of the problem. This method is called \textit{flow equation holography}.\cite{kehrein2017flow} It resembles the renormalization group (RG) flow approach and it has already been studied for the case of free fermions.\cite{kehrein2017flow} In this work, we go one step further by including interactions in the system. In particular, we study the case of weakly interacting fermions, on the example of the Hubbard model, and calculate the correction to the min-entanglement entropy with respect to free fermions (area law). Specifically, we are interested in knowing how the leading term in the expression for the min-entanglement entropy scales with the size of the entangled sub-region. 

\begin{figure}[h!]
\centering
  \includegraphics[width=0.45\textwidth]{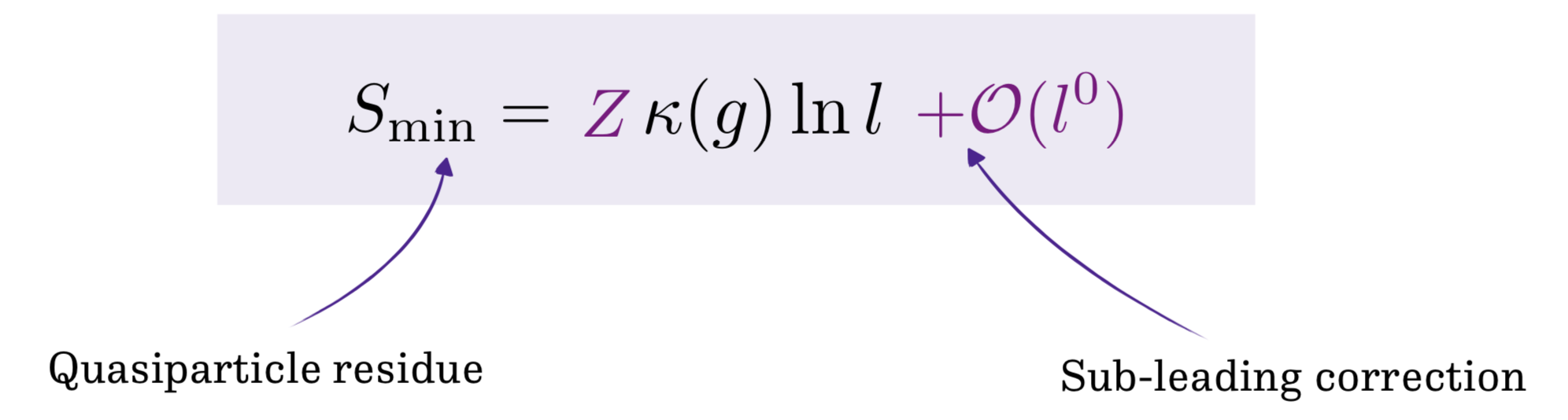}
  \caption{The main result of this work: expression of the min entanglement entropy for the system with \textit{weak local interactions}. The result is valid up to the order of $U^2$ in the interaction strength. $\kappa(g)$ is the pre-factor of the leading logarithmic term that corresponds to the free fermions in one dimension calculated in appendix \ref{appendix:free}, $\kappa(g) = g^2\dfrac{4}{\pi^2}\left(1 - \dfrac{\mu^2}{4}\right)$. It depends on the weak link parameter $g$ and the translationally invariant case is restored by putting $g=1$.}
  \label{fig:result}
\end{figure}

Explicit calculations for the one- and two-dimensional case (and similarly for higher dimensions) show that the leading correction to the entanglement entropy of non-interacting fermions up to $U^2$ in the interaction strength does not change the scaling, but only affects the pre-factor of the leading logarithmic term multiplying it by the quasiparticle residue. A term sub-leading to the area law is also present. It is proportional to $U^2$ in the interaction strength and scales linearly with the system size (see Fig. \ref{fig:result}).

The structure of this paper is following: in section I we give a brief overview of some basic concepts and techniques that are needed in order to understand the paper, such as entanglement entropy, its possible measures and scalings, flow equation and flow equation holography approaches. The latter is the main tool used in this paper. Section II is dedicated to the main part of this work, namely the calculation of the correction to the min entanglement entropy due to the presence of weak interactions in the system. Only the most important theoretical ideas and conceptual steps are present in the main text, whereas all the technicalities and auxiliary calculations can be found in the appendices.

\section{Basic concepts}
\textit{Min-entanglement entropy---}
Various measures of the entanglement entropy exist depending on the partition of the system (here, we consider only bi-partite entanglement entropies) and on the information it carries.\cite{vedral1998entanglement, vedral1997quantifying, horodecki2001entanglement, eisert1999comparison}. A generally accepted and widely used entanglement measure for bi-partite systems is the quantum Renyi entropy\cite{eisert2008area, horodecki2009quantum}
\begin{equation}\label{eq:renyi_entropy}
S_\alpha = \frac{1}{1-\alpha}\log \left(\Tr \hat{\rho}^\alpha\right),
\end{equation} 
where $\hat{\rho}$ is the density matrix of the bi-partite system and the trace is performed over the subsystem. This entanglement measure parametrises the whole family of measures depending on the order parameter $\alpha\geq 0$ and $\alpha\neq 1$. 

The limit $\alpha \rightarrow \infty$ defines a very informative and useful entanglement measure which is called \textit{min-entropy}.\cite{konig2009operational, smith2011quantifying, espinoza2013min} It defines a lower bound on entanglement entropy that is equal to the maximal eigenvalue of the reduced density matrix. This kind of entanglement entropy has a deeper physical meaning, namely, it represents the informational content that is most likely to come out during the random experiment.\cite{konig2009operational} There is also an interpretation of the min entanglement entropy as a single-copy distillation entropy.\cite{horodecki2001entanglement}

Scaling of the entanglement entropy with the subsystem size can characterize the macroscopical fundamental properties of the system such as criticality, topological properties, the measure of quantum chaos in the system, etc. The most typical scaling is a so-called \textit{volume law}, which is inherent to the thermal states, excited states and quantum chaotical systems. A much less common case is the \textit{area law} scaling which is typical of ground states. Usually, it is found in the literature that the entanglement entropy satisfies an area law if $S(\rho_I) = \mathcal{O}(s(I))$, where $s(I)$ is the length of the boundary of the subsystem $I$. \cite{eisert2008area}

One can expect that the addition of weak interactions to the system will not qualitatively influence the leading order of the scaling law. It can, however, introduce a pre-factor and/or give rise to the sub-leading terms with a different scaling. The current work aims to check this conjecture.  

\textit{Flow equation---}
The main focus of this paper is the application of the flow equation holography method\cite{kehrein2017flow} to the calculation of the min entanglement entropy of a weakly interacting system. The idea behing this method resembles the so-called flow equation method\cite{wegner1994flow,glazek1993renormalization}, which is an analytical tool for the Hamiltonian diagonalization. This technique allows us to find the approximate set of eigenmodes in a perturbative way. 

\begin{figure}[htb!]
\centering
  \includegraphics[width=0.4\textwidth]{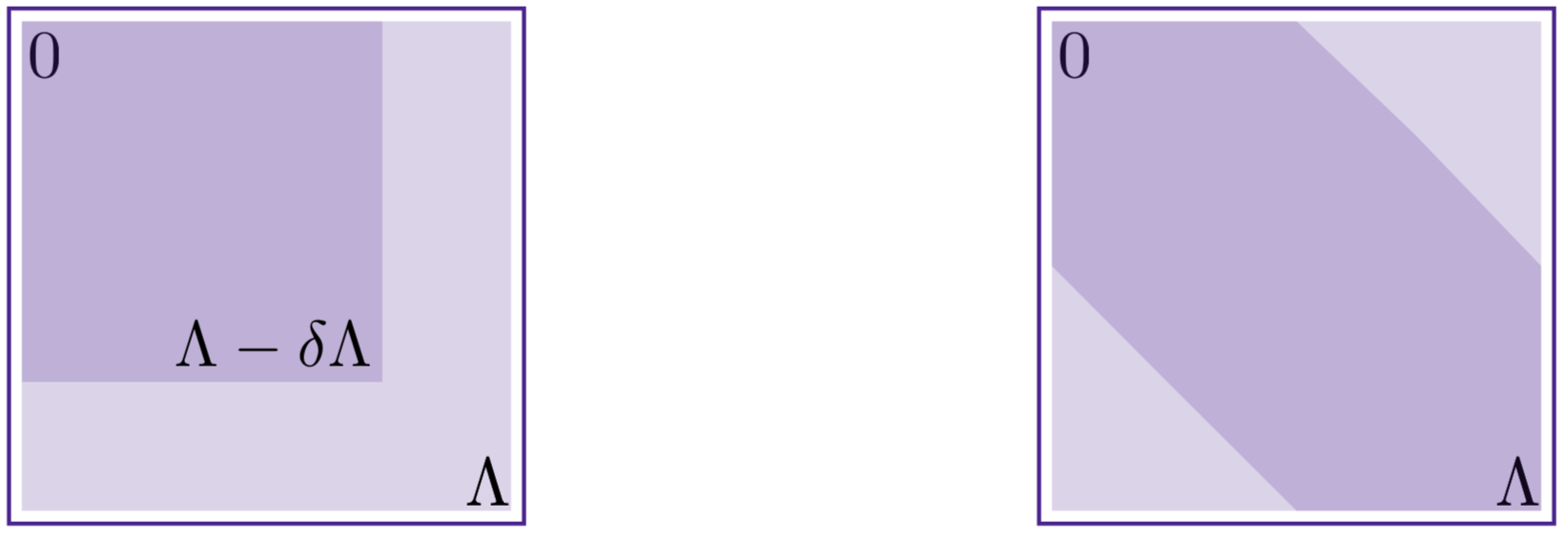}
  \caption{Schematic representation of the conventional scaling methods (left) and of the flow equation approach (right). In the former, one integrates out high energy degrees of freedom and arrives at an effective Hamiltonian that describes the low-energy sector. In the latter, one gets rid of the off-diagonal elements that correspond to high energy \textit{differences} with the help of a unitary transformation.}
  \label{fig:rg_vs_flow}
\end{figure}

The flow equation method is based on the energy scale separation concept, where we iteratively reduce the ultraviolet (UV)-cutoff $\Lambda_{\text{RG}}$ in \textit{energy differences}, or frequencies.\cite{kehrein2007flow, wegner1994flow, glazek1993renormalization} These energies correspond to the off-diagonal matrix elements of $\hat{H}$. In contrast to the conventional scaling methods, here all the energy scales are preserved (see Fig. \ref{fig:rg_vs_flow}). 

The task then boils down to finding the generator that will diagonalize the Hamiltonian during the flow, \textit{i.e.,} the generator that will transform the Hamiltonian in a way to satisfy 
\begin{equation}\label{eq:flow_generator_condition}\frac{d}{dB} \Tr \left( \hat{H}_{\text{int}}^2\right)\leq 0.
\end{equation}

This can be solved by the Jacobi method giving the following generator
\begin{equation}\label{eq:flow_generator}
\hat{\eta}(B) = \left[\hat{H}_0(B), \hat{H}_{\text{int}}(B) \right],
\end{equation}
where $\hat{H}_0(B)$ denotes the diagonal part of the Hamiltonian during the flow, and $ \hat{H}_{\text{int}}(B) $ the interaction part that we want to eliminate. Ideally, in the limit of infinitely long flow, we get diagonal Hamiltonian. In practice, however, there can be some difficulties to reach the final diagonal matrix. The reason is that the flow becomes much harder to calculate when the states are almost degenerate.\cite{moeckel2008interaction}
Typically, the expression can diverge starting from some moment in the flow, $B^*$. 
In this case, we say that the Hamiltonian of such a system can be diagonalized only to \textit{some} extent. The parameter $B^*$ also depends on the energy of the quasi-particles, \textit{e.g.}, for some systems the diagonalization can be exact only on the Fermi surface, while away from it one gets degeneracies which indicate non-perturbative effects 
This will be discussed in more detail in the subsection called ``Disentangling flow" of section \ref{sec:min_ent}.

\textit{Flow equation holography---}
When it comes to an explicit calculation of the entanglement entropy, the situation is very similar to the Hamiltonian diagonalization: very little can be done analytically even for non-interacting systems.\cite{eisert2008area, peschel2012special}
However, the scientific community keeps coming up with new clever ways to bypass complications and bottlenecks, and extract the information needed.\cite{peschel2012special}
Recently, a new powerful method called \textit{flow equation holography} was introduced \cite{kehrein2017flow} that allows for a non-perturbative, analytical calculation of the min entanglement
entropy of the general many-body system.
This method provides a systematic procedure which connects the entanglement properties of eigenstates of a generic quantum many-body Hamiltonian to a disentangling flow in an emergent RG-like dimension. The method is conceptually similar to the flow equation approach and uses the same RG-like flow, here, however, in a very different manner.

\begin{figure}[h!]
\centering
  \includegraphics[width=0.25\textwidth]{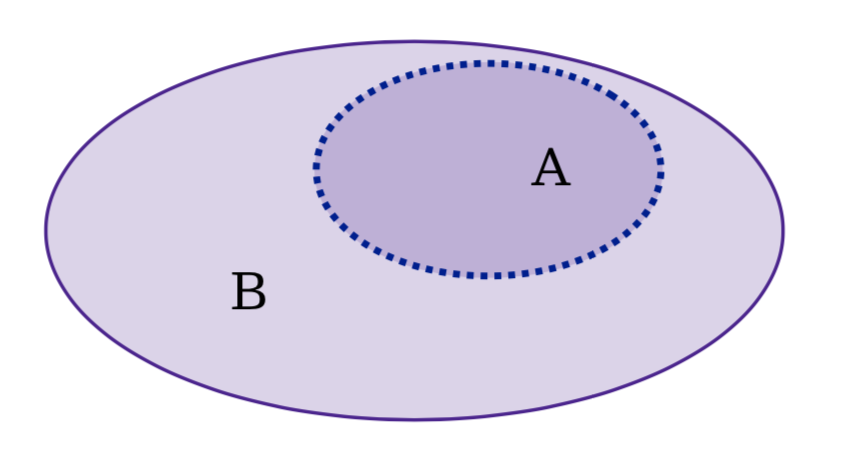}
  \caption{Schematic representation of a bi-partite system with a weak link, that is represented by a dotted line.}
  \label{fig:system}
\end{figure}

To be more precise, we divide our Hilbert space $\mathcal{H}$ into two sub-spaces $\mathcal{A}$ and $\mathcal{B}$ which are, generally, entangled between each other (see Fig. \ref{fig:system}). Correspondingly, the Hamiltonian of the whole system is 
\begin{equation}\label{eq:Hamiltonian_partition}\hat{H} = \hat{H}_{\mathcal{A}}\bigotimes\mathds{1}_{\mathcal{B}}+ \mathds{1}_{\mathcal{A}}\bigotimes\hat{H}_{\mathcal{B}} + g\hat{H}_{\text{ent}}.
\end{equation} 

 The first two terms act in a non-trivial way only on their corresponding subsystems, while the part $\hat{H}_{\text{ent}}$ is responsible for the entanglement between two subsystems.
 The coefficient $g\ll1$ is called \textit{weak link parameter} that couples two subsystems.

Now, in the very same way as we got rid of the non-diagonal part of the Hamiltonian in the flow equation approach, we use the unitary flow to eliminate the entangling part of the Hamiltonian $\hat{H}_{\text{ent}}$ here. In the end, when the flow parameter is sent to infinity $B \rightarrow \infty$, our Hilbert space is a direct sum of two sub-spaces that correspond to two subsystems $\mathcal{H} = \mathcal{H}_{\mathcal{A}\mathcal{B}} =  \mathcal{H}_{\mathcal{A}}\bigoplus\mathcal{H}_{\mathcal{A}} $ and the subsystems are completely decoupled. By building the analogy between the flow equation and flow holography methods, one can easily deduce the flow generator in this case 
\begin{equation}\label{eq:entangling_flow_generator}
\hat{\eta}(B) = \left[ \hat{H}(B), g\hat{H}_{\text{ent}}(B) \right].
\end{equation}

\begin{figure}[h!]
\centering
  \includegraphics[width=0.5\textwidth]{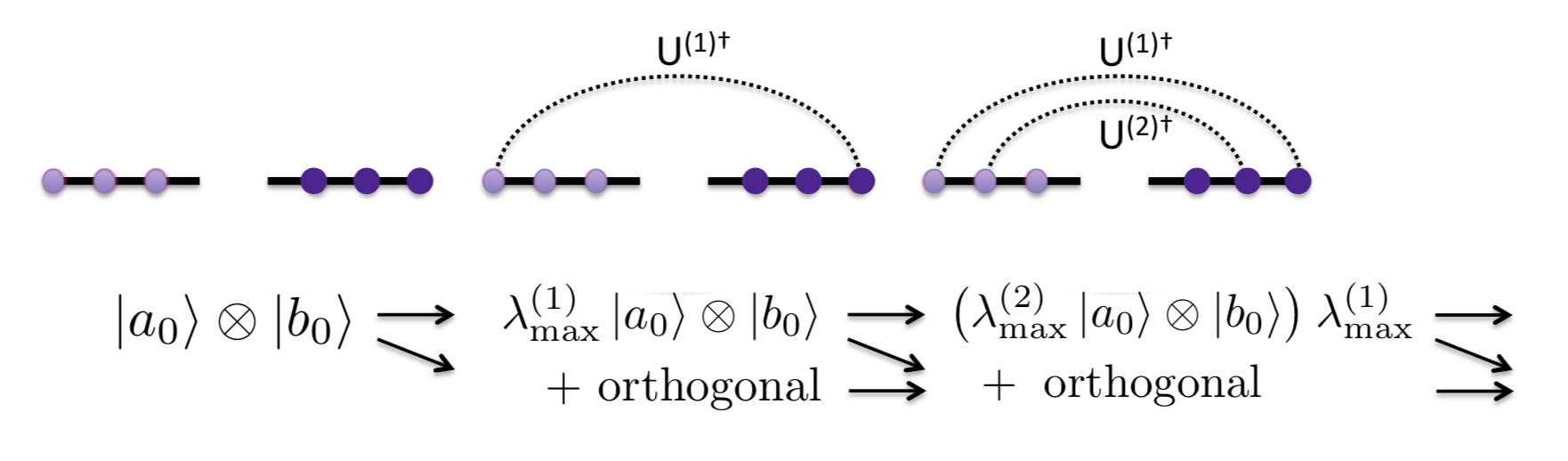}
  \caption{Schematic evolution of the state during the \textit{entangling} flow starting from the product state. The scheme tracks the largest Schmidt coefficient of the state. The figure is taken from \cite{kehrein2017flow}.}
 \label{fig:factorization}
\end{figure}

The structure of the generator implies that the flow is iterative in its nature and each step is a separate unitary transformation that eliminates the corresponding off-diagonal element labelled by some energy difference.
Which means that the transformation acts on sites that are further and further apart from the boundary between regions $\mathcal{A}$ and $\mathcal{B}$ as the flow parameter increases. A schematic representation of the process can be seen in Fig. \ref{fig:factorization}, where, however, the flow is reversed.

By using this crucial property, one can derive the analytical expression of the min-entropy \cite{kehrein2017flow}
\begin{equation}\label{eq:main}
S_{\text{min}} = -4\int_{\frac{1}{\Lambda^2}}^{\infty} BdB \bra{\psi}\hat{\eta}^2(B)\ket{\psi} + \mathcal{O}(g^3).
\end{equation}
Here, $\Lambda$ corresponds to the ultra-violet cut-off, which reflects the fact that we consider the system on a lattice with a finite lattice constant $a$ and $\ket{\psi}$ is a ground state of the decoupled system. This is needed in order to get rid of the divergences that arise at $B=0$. 

\section{Min entanglement entropy calculation}
\label{sec:min_ent}
\textit{Model---}
Let us consider the bi-partite system ($\mathcal{A}$ and $\mathcal{B}$, see Fig. \ref{fig:system}) that is filled with locally interacting fermions. These two sub-regions are weakly coupled by spin-symmetric hopping terms. For the sake of simplicity, the interaction is turned on only in the region $\mathcal{B}$. This does not influence the final result qualitatively and affects only the numerical pre-factor. The Hamiltonian reads
\begin{equation*}\label{eq:hubbard}
\hat{H}_\mathcal{A} = -\sum_i \sum_{\sigma} \hat{a}^{\dagger}_{\sigma,i} \hat{a}_{\sigma,i+1} + U\sum_i \hat{n}^{\uparrow}_i \hat{n}^{\downarrow}_i + \text{h.c.}
\end{equation*}
\begin{equation*}
\hat{H}_\mathcal{B} = -\sum_i \sum_{\sigma} \hat{b}^{\dagger}_{\sigma,i} \hat{b}_{\sigma,i+1} + \text{h.c.}
\end{equation*}
\begin{equation}
\hat{H}_{\text{ent}} = -g \sum_{\sigma} \hat{a}^{\dagger}_{\sigma,0} \hat{b}_{\sigma,1} + \text{h.c.}
\end{equation}

For what concerns the non-interacting part of the Hamiltonian, this would lead to the same result as for free fermions (see appendix \ref{appendix:free}). The presence of the interacting term in the Hamiltonian $\mathcal{A}$ creates a different situation where the entanglement of the subsystem is defined by the competition between the kinetic part, that would lead to growing entanglement entropy, and the interacting one where  the entanglement is zero because the state is an insulating one.

\textit{Treatement of the interactions using the flow equation technique---}
Before we start the consideration of the disentangling flow, we first need to diagonalize the interacting Hamiltonian of the subsystem $\mathcal{A}$. In this way, we will obtain the eigenmodes which are needed for the considerations that will follow, such as the calculation of the expectation value of the flow generator for the disentangling flow. The detailed calculations for this section can be found in appendix \ref{appendix:diag}, here, we only sketch the main steps to give a general idea.

Since we assumed the subsystem $\mathcal{B}$ to be non-interacting, the corresponding fermionic degrees of freedom can be described using an eigenbasis composed of plane waves.
The dispersion relation is then the usual  $\epsilon_k = -2\cos k$, where $k$ is the wave vector. The part of the Hamiltonian that acts on the subsystem $\mathcal{B}$ can be then written as
\begin{equation}
\hat{H}_\mathcal{B} = -\sum_{\sigma}\sum_k \e_k\bd_{\sigma,k}\bn_{\sigma,k} .
\end{equation}

If we naively apply the same transformation to the Fermi-Hubbard Hamiltonian of subsystem $\mathcal{A}$, the presence of local interactions in the real space would make our Hamiltonian non-diagonal in the energy basis
\begin{equation}
\hat{H}_\mathcal{A} = -\sum_{\sigma}\sum_k \e_k\ad_{\sigma,k}\an_{\sigma,k}+ U\sum_{\alpha,\beta,\gamma,\delta}\adup_{\alpha}\anup_{\beta}\addo_{\gamma}\ando_{\delta} \delta^{\alpha+\gamma}_{\beta + \delta}.
\end{equation}

Therefore, one has to diagonalize  $\hat{H}_\mathcal{A}$ with the flow equation method \cite{wegner1994flow}.
 Fermi gas in the thermodynamic limit represents a many-particle problem with an infinite number of degrees of freedom. This implies that many different energy scales contribute to the Hamiltonian. It is obvious that the interaction generates occupation in many different excited eigenstates of the interacting Hamiltonian. Therefore we implement the diagonalizing transformation such that a controlled treatment of different energy scales is possible \cite{moeckel2008interaction}.
  This is exactly what the flow equation method can offer us.

\begin{figure}[h]
\centering
  \includegraphics[width=0.5\textwidth]{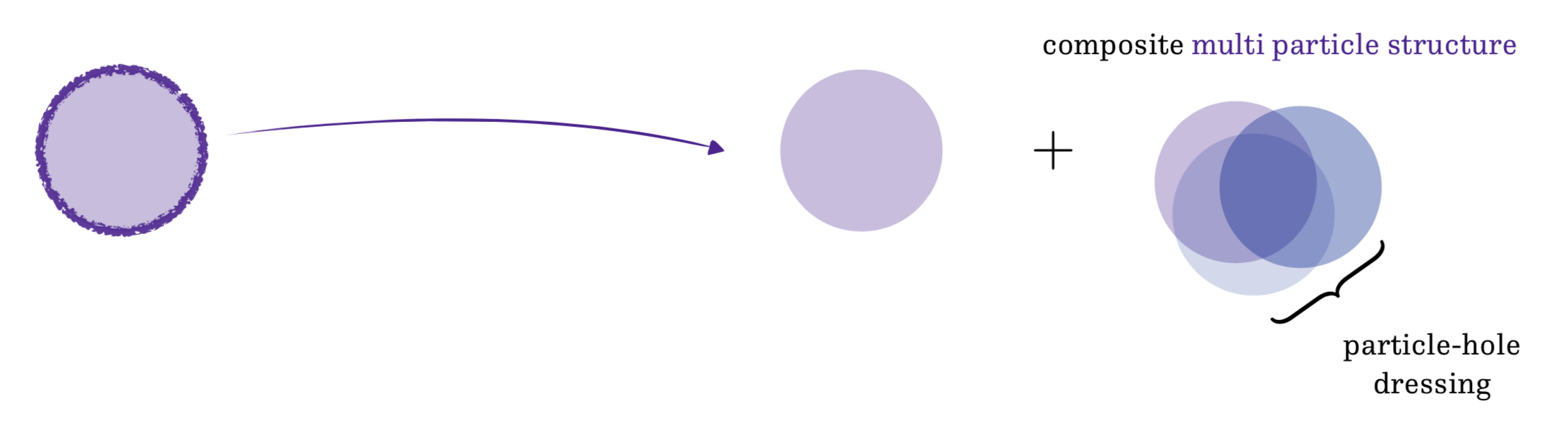}
  \caption{The schematic representation of the multi-particle structure that the quasi-particle acquire during the diagonalization flow.}
  \label{fig:quasi}
\end{figure}

Before to look into the technical details, one can develop an intuitive picture of the process: during the flow one-particle creation and annihilation operators transform and acquire multi-particle coating (see Fig. \ref{fig:quasi}). Also, different spins mix between each other such that the particle with some spin $\sigma$ becomes a multi-particle excitation of different spins. Such excitation is called \textit{quasi-particle} and is denoted by new creation and annihilation operators ($\cn, \cd$ in the subsystem A). In the basis of these operators, our Hamiltonian has a diagonal form. However, in the same fashion as for the conventional RG-flow, not only the operational structure of the Hamiltonian changes, but also  the parameters in the Hamiltonian  become ``flowing", see Eq. \ref{eq:diagonal_H}. 

\vspace{0.3cm}

One can write down a physical ansatz for the quasparticles\cite{moeckel2008interaction}, where the functional dependence is deduced from the flow differential equations (see appendix \ref{appendix:diag}) and get the expression for the diagonalized Hamiltonian
\begin{widetext}
\begin{align}
\nonumber \hat{H} = -\sum_{\sigma}\sum_k \e_k\adup_{k,\sigma}\anup_{k,\sigma} -\sum_{\sigma}\sum_k \e_k\bdup_{k,\sigma}\bnup_{k,\sigma} &-\sum_{\sigma}\sum_{l,m}\dfrac{2}{N+1}\sin m \sin l \lb h_l(\infty)g_{l,m}(B)\ad_{l,\sigma} \bn_{m,\sigma} + \right.\\
&\left.+\sum_{\pp \qp p} M^l_{\pp\qp p}(\infty)g_{\pp\qp p,l,m}(B) :\ad_{\pp,\sigma}\ad_{\qp,-\sigma}\an_{p,-\sigma}:\bn_{m,\sigma} + h.c \rb
\end{align}
with
\begin{align}
 M^l_{\pp\qp p}(B) &= U\rho\int^{\infty}_{-\infty} dE\dfrac{ \delta(E - \e_\pp - \e_\qp + \e_p)}{(E-\e_l)} \lb 1 - e^{-B(E-\e_l)^2} \rb\\
 h_l(B) &= 1 - U^2\rho^3\int^{\infty}_{-\infty} dE\dfrac{(E-\e_F)^2}{(E-\e_l)^2} \lb 1 - e^{-B(E-\e_l)^2} \rb.
\end{align}
\end{widetext}

The explicit calculation is done in appendix \ref{appendix:diag}. It is important to emphasise that the flow function $h_l(B)$ is connected to the quasi-particle residue via $h_{k_F} = \sqrt{Z(B)}$ and depicts the coherent overlap of the physical fermion with the related interaction-free momentum mode (quasi-particle) of the current representation.

\textit{Disentangling flow---}
In the previous section, we considered the flow that diagonalized the Fermi Hubbard Hamiltonian. The corresponding flow parameter $B$ was sent up to infinity in order to reach perfect diagonalization. However, as was mentioned in the introduction, for some systems such procedure may diverge which means that the corresponding Hamiltonian can be diagonalized only up to a certain extent. In order to control all the divergencies, we will keep the final value of the previous diagonalising flow parameter at some fixed (big enough in order to assume that the Hamiltonian is diagonal) value $B^*$ and proceed to the next (disentangling) flow with the corresponding parameter $B$. Then, in the end, we will see whether we can send $B^*$ having all the expressions finite. And if we can not: what are the limitations for it.

In the end, we want to see how the interactions influence the entanglement entropy, therefore, we are keeping track of all the difference with the same calculations for free fermions.

Keeping in mind the commutation relations for the fermions one can work out the generator of the flow. This will, however, only give the operational form of it. In order to find the functional dependence for the  $g_{l,m}(B), g_{\pp\qp p,l,m}(B)$ on should solve the differential equation\cite{kehrein2017flow}
\begin{equation}\label{eq:diff_eq_flow}
d\hat{H}/dB = -\left[\hat{H}(B), \hat{\eta}(B)\right]
\end{equation} 
obtaining
\begin{align}\label{eq:solutions_g}
 g_{l,m}(B) &= g e^{-B(\e_l- \e_m)^2}\\
 g_{\pp\qp p,l,m}(B) &= g e^{-B(\e_{\pp} +\e_{\qp}- \e_p-\e_m)^2}.
\end{align}
Plugging the solutions \ref{eq:solutions_g} in the Eq. \ref{eq:flow_generator} we obtain the full form of the flow generator

\begin{widetext}
\begin{align}\label{eq:flow_fin}
\nonumber& \hat{\eta}(B) = \sum_{\sigma}\sum_{m,l}\dfrac{2g}{N+1}\sin m \sin l  \left[h_le^{-B(\e_l- \e_m)^2} \lb \ad_{l,\sigma}\bn_{m,\sigma} - \an_{l,\sigma}\bd_{m,\sigma}\rb(\e_l - \e_m) + \right.\\
+ \sum_{\pp \qp p} &\left. M^l_{\pp\qp p}e^{-B(\e_{\pp} +\e_{\qp}- \e_p-\e_m)^2}\lb :\ad_{\pp,\sigma}\ad_{\qp,-\sigma}\an_{p,-\sigma}\bn_{m,\sigma}:-:\an_{\pp,\sigma}\ad_{p,-\sigma}\an_{\qp,-\sigma} \bd_{m,\sigma}: \rb(\e_{\pp} +\e_{\qp}- \e_p-\e_m)  \right].
\end{align}
\end{widetext}

Flow generator that we obtained enters the expression for the entanglement entropy  in the form of the following expectation value: $\bra{\psi_0}\hat{\eta}^2\ket{\psi_0}$. Term  $\sim h\times h$ will give the same contribution as in the case for the free fermions but multiplied by the quasi-particle residue. Cross-terms (the ones $\sim h\times M$) will not contribute to the expectation value at all as there are the unequal amount of creation and annihilation operators in them and therefore the expectation value vanishes.

The only new and ``non-trivial" contribution to $\langle\hat{\eta}^2\rangle$ can arise from the $ M\times M$ term, which is $\sim U^2$, which one can work out with the help of the Wick theorem. By considering this term, we will calculate the correction $\delta S = S_{\text{interacting}} - ZS_{\text{free}}.$
The scaling of this additional term with the subsystem size is not known and is of a fundamental interest. By inserting the corresponding part of the $\hat{\eta}^2$ in Eq. \ref{eq:main} and performing the integration (see appendix \ref{appendix:int}) we obtain

\begin{widetext}
\begin{align}\label{eq:entropy_correction}
 \delta S &=   \dfrac{6g^2U^2 \rho}{\pi^2\sqrt{2\pi}} \lb\dfrac{\Lambda}{l} - \Lambda\rb\int^{\infty}_{-\infty} d \tilde{\e_l} \int_{0}^{\infty} d\tilde{E}  \dfrac{\tilde{E}^2}{(\tilde{E}+\tilde{\e_l})^2}\text{erfc}(\tilde{E}).
\end{align}
\end{widetext}

\textit{Final remarks---}
If one naively looks at the Eq.~\ref{eq:entropy_correction}, they will immediately see that the expression has a pole of the second order at $-\tilde{E} = \tilde{\e}_l$. The origin of this is hidden in the first diagonalization flow procedure. As was mentioned before, for some energies of the quasi-particle we can not diagonalize the Hamiltonian precisely, but only make it ``nearly" diagonal. Therefore, in order to avoid divergences one has to try the same procedure, but cutting the diagonalization flow of $\hat{H}_A$ at some value $B^*$. Presumably, this threshold $B^*$ can depend on the energy $\e$.  The further we go from the Fermi energy, the smaller is the lifetime of the quasi-particle $\tau \sim (\e-\e_F)^{-2}$. That is why those quasi-particles with a very small lifetime just do not manage to penetrate far enough into the other subsystem in order to entangle because they decay before doing it.

In order to perform the same calculation keeping all the divergences under control, we act iteratively: diagonalize the Hamiltonian up to some extent, then entangle only up to some energies from the Fermi level, then proceed to better diagonalization, then entangle further and so on. Indeed, if one re-considers all the calculations keeping $B^*$ finite, one would see that additional pre-factor that arises cancels the pole when $\tilde{E} \rightarrow -\tilde{\varepsilon}_l$. The regularised result has the same scaling and we leave the technical details of the analysis in appendix \ref{appendix:int}.

\bibliography{paperbiblio}

\begin{widetext}
\appendix
\section{Flow equation holography for free fermions}\label{appendix:free}
Consider one--dimensional free fermions on a lattice. We divide it into two subsystems $\mathcal{A}$ and $\mathcal{B}$ on the $0^{th}$ lattice site with $N_{\mathcal{A}}$  and $N_{\mathcal{B}}$ sites respectively (for simplicity we take $N_{\mathcal{A}} = N_{\mathcal{B}}$). The weak coupling $g$ between the two subsystems which is realised via the hopping term on the $0^{th}$ cite (junction) couples these subsystems. One can restore the trnslationaly invariant case by putting $g=1$. The Hamiltonian reads
\begin{align}
\nonumber \hat{H} & = \hat{H}_{\mathcal{A}}+ \hat{H}_{\mathcal{B}} + \hat{H}_{\text{ent}} = -t\sum_{n=N}^{2}(\ad_i\an_{i+1} + h.c)-t\sum_{n=N}^{2}(\bd_i\bn_{i+1} + h.c) - g(\ad_1\bn_1+ h.c).
\end{align}

By performing the entangling flow~\cite{kehrein2017flow}, these subsystems will entangle up to some length $\textit{l}$. For this, we calculate the min-entanglement entropy following the original work~\cite{kehrein2017flow}. In the system, where $t = 1$ the flow generator is

\begin{equation}\label{eq:generator_free}
\hat{\eta}(B) = \sum_{l,m} \dfrac{2g e^{-B(\e_l- \e_m)^2}\sin m\sin l}{N+1} \lb \ad_{l}\bn_{m} - \an_{l}\bd_{m} \rb   \lb \e_l - \e_m\rb,
\end{equation}
where $B$ is the flow parameter. Substituting the expression for the flow parameter~\ref{eq:generator_free} into the Eq.~\ref{eq:main} we obtain the answer for the min entanglement entropy depending on the weak-link parameter $g$
\begin{align}\label{eq:result_free}
  S_{\text{min}}=g^2\dfrac{4}{\pi^2}\left(1 - \dfrac{\mu^2}{4}\right)\ln l, 
\end{align}
where $l$ is the length of the subsystem and $\mu$ is chemical potential. The answer in a perfect agreement with the previous works\cite{kehrein2017flow} and represents a logarithmic correction to the Area Law as should be for the critical systems\cite{calabrese2004entanglement}.
 For the translation-invariant case $g=1$ the behaviour is logarithmic in $l$ with the universal pre-factor independent of the filling\cite{calabrese2004entanglement}.
 
It can be shown that the two-dimensional free fermions are equivalent to the set of one-dimensional free fermions gases with momenta-dependent chemical potential $\mu(k_y)$ which makes the answer for the min entanglement look as follows

\begin{align}
S_{\text{min}}&= \dfrac{g^2N_y}{2\pi^3}\ln l \lb (2-\mu^2)\arcsin (\frac{\mu}{2}-1) + \sqrt{\mu - \frac{\mu^2}{4}}(6\mu -4) \rb .
\end{align}
\section{Diagonalization of the Fermi Hubbard Hamiltonian by means of the flow equation}
\label{appendix:diag}

The subsystem $\mathcal{A}$ with local interactions can be represented by the Fermi Hubbard Hamiltonian which in the basis of plane waves reads
\begin{equation}
\hat{H}_{\mathcal{A}} = -\sum_{\sigma}\sum_k \e_k\ad_{\sigma,k}\an_{\sigma,k}+ U\sum_{\alpha,\beta,\gamma,\delta}\adup_{\alpha}\anup_{\beta}\addo_{\gamma}\ando_{\delta} \delta^{\alpha+\gamma}_{\beta + \delta}.
\end{equation}
Using the standard normal ordering we can rewrite the  interacting part as
\begin{equation}\label{eq:normal}
\adup_{\alpha}\anup_{\beta}\addo_{\gamma}\ando_{\delta} = :\adup_{\alpha}\anup_{\beta}\addo_{\gamma}\ando_{\delta}: + n_{\gamma}\delta^\gamma_\delta :\adup_{\alpha}\anup_{\beta}: + n_{\alpha}\delta^\alpha\beta:\addo_{\gamma}\ando_{\delta}: - n_{\gamma}n_{\alpha}\delta^\gamma_\delta\delta^\alpha\beta.
\end{equation}
The last term of the Eq.~\ref{eq:normal} is a constant off-set which can be ignored. The second and the third terms give the contribution to the energy $\e_k$ of order $U$. The perturbative series we are going to see in some lines makes clear that this contribution will be subdominant in the approximation we are going to consider therefore we can neglect these terms. 

Hamiltonian $\hat{H}_\mathcal{A}$ is diagonalised by the means of flow equation method~\cite{kehrein2007flow}. The new basis corresponds to the complex multi-particle states (\textit{quasi-particles} $\cn, \cd$). Not only the operational structure of the Hamiltonian will change, but also the parameters in the Hamiltonian will become "flowing". The final Hamiltonian reads
\begin{align}\label{eq:diagonal_H}
\hat{H} = -\sum_{\sigma}\sum_k \e_k(\infty):\cd_{\sigma,k}\cn_{\sigma,k}:& -\sum_{\sigma}\sum_k \e_k:\bd_{\sigma,k}\bn_{\sigma,k}: -\sum_{\sigma}\sum_{l,m}\dfrac{2}{N+1}\sin m\sin l g_{l,m}(\infty)\lb \cd_{\sigma,l}\bn_{\sigma,m} + h.c\rb.
\end{align}

The operational form of the quasiparticle creation and annihilation operators $\cn, \cd$  can be guessed and set as an ansatz to the differential equations of the flow in order to find the functional dependence.\cite{moeckel2008interaction}
 This can be done as far as Eq. \ref{eq:differential_eq}.  is full-filled not only by the generator of the flow but also by any Hermitian operator
\begin{equation}\label{eq:differential_eq}
\dfrac{d\hat{\mathcal{O}}(B)}{dB} = \left[\hat{\eta}(B),\hat{\mathcal{O}}(B)\right],
\end{equation}
where $B$ is the flow parameter. The ansatz for the physical fermion includes the lowest order corrections  and takes into account spin and momentum conservations (see Eq. \ref{eq:ansatz}). Note, that the particle that had some definite spin $\sigma$ in the system without interactions is now ``coated" by particles of different spins including $-\sigma$.
 
\begin{align}\label{eq:ansatz}
\nonumber \cd_{l,\sigma} &= h^{\sigma}_l(B)\ad_{l,\sigma} + \sum_{\pp \qp p} M^{l,\sigma}_{\pp\qp p}(B) :\ad_{\pp,\sigma}\ad_{\qp,-\sigma}\an_{p,-\sigma}:+ \sum_{\pp \qp p} M^{l,-\sigma}_{\pp\qp p}(B) :\ad_{\pp,\sigma}\ad_{\qp,\sigma}\an_{p,\sigma}:\\
\cn_{l,\sigma} &= h^{\sigma}_l(B)\an_{l,\sigma} + \sum_{\pp \qp p} M^{l,\sigma}_{\pp\qp p}(B) :\an_{\pp,\sigma}\ad_{p,-\sigma}\an_{\qp,-\sigma}:+ \sum_{\pp \qp p} M^{l,-\sigma}_{\pp\qp p}(B) :\an_{\pp,\sigma}\ad_{p,\sigma}\an_{\qp,\sigma}:.
\end{align}
Here $h^{\sigma}_l(B), M^{l,\sigma}_{\pp\qp p}(B) $ are the flowing functions of the observable $\cd_{l,\sigma}$ and can be deduced from the differential equations \ref{eq:differential_eq}. The physical meaning of $h^{\sigma}_l(B)$ is simple: it is connected to the quasi-particle residue via $h_{k_F} = \sqrt{Z(B)}$ and depicts the coherent overlap of the physical fermion with the related interaction-free momentum mode (quasi-particle) of the current representation. The function $M^{l,\sigma}_{\pp\qp p}(B)$ represents the incoherent background in the spectral function of an interacting system~\cite{moeckel2008interaction}. The differential equations on these functions for the particular case up spin up are
\begin{align}\label{eq:differential_for_constants}
\dfrac{\partial \hup_l(B)}{\partial B} &= U \sum_{p\qp q}\Delta \e_{l,p\qp q}e^{B(\Delta \e_{ l,p\qp q})^2} Q_{p q\qp}M^{l,\uparrow}_{p q\qp}(B) \\
\dfrac{\partial M^{l,\downarrow}_{\alpha\beta\gamma}(B) }{\partial B} &= -U \sum_{\qp q}\left[ n_{\qp} - n_q\right]\Delta \e_{\qp q\alpha \gamma}e^{B(\Delta \e_{ \qp q\alpha \gamma})^2} M^{l,\uparrow}_{\beta \qp q}(B)\\
\nonumber \dfrac{\partial M^{l,\uparrow}_{\alpha\beta\gamma}(B) }{\partial B} &= U \sum_p \hup_p(B)\Delta \e_{\alpha \gamma\beta p}e^{B(\Delta \e_{ \alpha \gamma\beta p})^2}   +\\
\nonumber&+U \sum_{p \qp}\left[ n_{p} - n_{\qp}\right]\Delta \e_{\qp p\alpha \gamma}e^{B(\Delta \e_{\qp p\alpha \gamma})^2} M^{l,\uparrow}_{p \beta \qp }(B)\\
\nonumber&+U \sum_{p q}\left[1+ n_{q} - n_{p}\right]\Delta \e_{\alpha p \beta q}e^{B(\Delta \e_{\alpha p \beta q})^2} M^{l,\uparrow}_{p q \gamma }(B)\\
&+U \sum_{\pp p}\left[ n_{\pp} - n_{p}\right]\Delta \e_{\pp p \beta \gamma}e^{B(\Delta \e_{\pp p \beta \gamma})^2}\left[  M^{l,\downarrow}_{p \gamma \pp}(B) -M^{l,\downarrow}_{\alpha p \pp}(B)\right].
\end{align}

As far as we are interested only in the leading order in $U$, it will be easy to truncate the chain of differential equations \ref{eq:differential_for_constants} and they will not be coupled any more. However, the solution will depend on its initial conditions. Since the differential equations,  \ref{eq:differential_for_constants} are linear, a solution for general initial conditions can be achieved as a linear superposition of solutions for independent initial configurations. There are two possible cases:
\begin{enumerate}
\item \textit{Fully coherent initialization} of one fermion in the momentum mode $k$:

$\hup (B=0) = \delta^i_k$ and $M(B=0) = 0$ for all possible indices.
\item \textit{Fully incoherent initialization}  in the dressing state $\pp \qp p, \uparrow$:

$\hup (B=0) = 0$, $M^{\uparrow}_{\alpha \beta \gamma}(B=0) = \delta^{\pp}_{\alpha} \delta^{\qp}_{\beta} \delta^p_\gamma$ and $M^{\downarrow}(B=0) = 0$  for all possible indices.
\end{enumerate}
We will only consider the first case of fully coherent initialization here and the other case is left for further studies. Now we can iteratively track the action of the differential flow equations at the onset of the flow. The first iteration is defined by the initial conditions. Then in the next iteration, keeping only terms up to order of $U$ we see that $M^{\uparrow}$ is defined only by the first term plugging in the initial condition for $h$. This implies that $M \sim U$ and this is the only term we will keep hence all the other will be of a higher order
\begin{equation}
\dfrac{\partial M^{l,\uparrow}_{\alpha\beta\gamma}(B) }{\partial B} = U \sum_p \delta^p_l\Delta \e_{\alpha \gamma\beta p}e^{B(\Delta \e_{ \alpha \gamma\beta p})^2}  .
\end{equation}
This equation can be easily solved and the solution should be plugged in the equation for $h$
\begin{equation}
\dfrac{\partial \hup_l(B)}{\partial B} = U \sum_{p\qp q}\Delta \e_{l,p\qp q}e^{B(\Delta \e_{ l,p\qp q})^2} Q_{p q\qp}M^{l,\uparrow}_{p q\qp}(B).
\end{equation}

We also see that term with $M^{\downarrow}$ will not contribute to the leading order at all as far as it only has terms proportional to $M\times U$ in its equations. This can be also understood from the Pauli principle: it is much favourable for electrons to occupy states with different spin directions when it is possible. The same rule is used for the well-known Madelung rules 
 All this means that the solution will be \textit{at least} of the order $U^2$ in the interaction strength. In the end, after all the simplifications we have the following system of equations

\begin{align}
\dfrac{\partial h^{\sigma}_l(B)}{\partial B} &= U\sum_{\pp\qp p} M^{l,\sigma}_{\pp\qp p}(B)\Delta_{l\pp p\qp} e^{-B \Delta^2_{l\pp p\qp}}Q_{\pp p\qp}\\
\dfrac{\partial M^{l,\sigma}_{\pp\qp p}(B)}{\partial B} &= h^{\sigma}_l(B) U \Delta_{\pp p\qp l} e^{-B \Delta^2_{\pp p\qp l}}.
\end{align}

We should, however, note that all the considerations were made for the spin up. Nevertheless, the equations and the solutions are symmetric for both values of the spin
\begin{align}
 M^l_{\pp\qp p}(B) &= U\dfrac{1 - e^{-B \Delta^2_{\pp p\qp l}}}{\Delta_{\pp p \qp l}} \\
 h_l(B) &= 1 - U^2\sum_{\pp \qp p}\dfrac{1 - e^{-B \Delta^2_{\pp p\qp l}}}{2\Delta^2_{\pp p \qp l}}Q_{\pp p\qp},
\end{align}
where $\Delta_{\pp p \qp l} = \e_{\pp} -\e_p +\e_{\qp} - \e_l$. One can insert and additional integral
\begin{align}
 M^l_{\pp\qp p}(B) &= U\rho\int^{\infty}_{-\infty} dE\dfrac{ \delta(E - \e_\pp - \e_\qp + \e_p)}{(E-\e_l)} \lb 1 - e^{-B(E-\e_l)^2} \rb\\
 h_l(B) &= 1 - U^2\rho^3\int^{\infty}_{-\infty} dE\dfrac{(E-\e_F)^2}{(E-\e_l)^2} \lb 1 - e^{-B(E-\e_l)^2} \rb.
\end{align}
The energy eigenmodes of the diagonal Hamiltonian are hence
\begin{align}
\nonumber \cd_{l,\sigma} &= h^{\sigma}_l(\infty)\ad_{l,\sigma} + \sum_{\pp \qp p} M^{l,\sigma}_{\pp\qp p}(\infty) :\ad_{\pp,\sigma}\ad_{\qp,-\sigma}\an_{p,-\sigma}:\\
\cn_{l,\sigma} &= h^{\sigma}_l(\infty)\an_{l,\sigma} + \sum_{\pp \qp p} M^{l,\sigma}_{\pp\qp p}(\infty) :\an_{\pp,\sigma}\ad_{p,-\sigma}\an_{\qp,-\sigma}:.
\end{align}

\section{Evaluation of the integral~\ref{eq:main} for the correction to the min-entanglement entropy}
\label{appendix:int}
The integral that we obtain after substituting the flow parameter~\ref{eq:flow_fin} to the Eq.~\ref{eq:main} and leaving only the term quadratic in the flow function $M$ we obtain
\begin{align}
\nonumber \delta S =-\sum_{l,m}\sum_{\pp \qp p} 4\int_{\lb\frac{1}{\Lambda}\rb^2}^{\lb\frac{l}{\Lambda}\rb^2} BdB \dfrac{16 g^2}{(N+1)^2} &e^{-2B(\e_{\pp} +\e_{\qp}- \e_p-\e_m)^2}(\e_{\pp} +\e_{\qp}- \e_p-\e_m)^2  \dfrac{\delta^{l+p}_{\pp+\qp}U^2\sin^2 m\sin^2 l}{(\e_{\pp}-\e_p+\e_{\qp} - \e_l)^2} \times\\
&\times\lb n_{\pp}n_{\qp}(1-n_p)(1-n_m) +(1- n_{\pp})(1-n_{\qp})n_pn_m\rb.
\end{align}

It is  useful to notice, that $2\sin m = \sqrt{4 - 4\cos^2 m} =  \sqrt{4 - \e_m^2} = 1/\rho(\e_m)$, where $\rho(\e_m)$ is a density of states. One can go from the sum over the wave-vector indices to the integral over energies. Even though the integration limits correspond to the borders of the energy band, namely $\e_l = 2\cos l \in [-2,2]$, we can expand the region of the integration to $[-\infty, \infty]$ due to the properties of the density of states (zero outside the energy band).

We can also take the densities of states on the chemical potential as an approximation. This can be motivated by the fact that in the region of high flow parameter $B$ (the one we are interested in) high energy differences are suppressed by the factor $ e^{-2B\lb \e_l - \e_m \rb^2 } $. The only value where both energies are the closest is at the chemical potential. So we obtain

\begin{align}
 \nonumber \delta S =  -\dfrac{2 g^2U^2 \rho}{\pi^2} \int_{\lb\frac{1}{\Lambda}\rb^2}^{\lb\frac{l}{\Lambda}\rb^2}  BdB&\int_{-\infty}^{\infty} \int_{-\infty}^{\infty} \int_{-\infty}^{\infty} \int_{-\infty}^{\infty} d\e_l d\e_m   d\e_{\pp}  d\e_{\qp}\delta(E - \varepsilon_{\pp} - \varepsilon_{\qp}+\varepsilon_p) e^{-2B(E-\e_m)^2}  \dfrac{(E-\e_m)^2}{(E-\e_l)^2}\times\\
 &\times\lb n_{\pp}n_{\qp}(1-n_p)(1-n_m) +(1- n_{\pp})(1-n_{\qp})n_pn_m\rb,
\end{align}
where we inserted one more integration over $E$ by using the Dirac delta-function. One can  get rid of $(E - \e_m)^2$ by taking the derivative over $B$
\begin{align}
 \nonumber\delta S =  \dfrac{g^2U^2 \rho}{\pi^2} \int_{\lb\frac{1}{\Lambda}\rb^2}^{\lb\frac{l}{\Lambda}\rb^2} dB B&\dfrac{\partial}{\partial B}\int_{-\infty}^{\infty} \int_{-\infty}^{\infty} \int_{-\infty}^{\infty} \int_{-\infty}^{\infty} \int_{-\infty}^{\infty} d\e_l d\e_m  dE d\e_{\pp}d\e_{\qp} \dfrac{\delta(E - \varepsilon_{\pp} - \varepsilon_{\qp}+\varepsilon_p)}{(E-\e_l)^2}e^{-2B(E- \e_m)^2}\times\\
&\times\lb n_{\pp}n_{\qp}(1-n_p)(1-n_m) +(1- n_{\pp})(1-n_{\qp})n_pn_m\rb.
\end{align}

Let us now take care of the part where we have integration over $E,\e_{p}, \e_{\qp},\e_m$. We see that such an integral is non-zero only in 2 cases:
\begin{enumerate}
\item $\e_{\pp}<\e_F,\e_{\qp}<\e_F, \e_{p}>\e_F,\e_{m}>\e_F$. In this case $E = \e_{\pp}+\e_{\qp}-\e_{p} \in [-\infty,\e_F]$
\item $\e_{\pp}>\e_F,\e_{\qp}>\e_F, \e_{p}<\e_F,\e_{m}<\e_F$. In this case $E = \e_{\pp}+\e_{\qp}-\e_{p} \in [\e_F, \infty]$
\end{enumerate}

So we have 2 terms
\begin{align}
\nonumber I = \int_{\e_F}^{\infty} d\e_m  \int_{-\infty}^{\e_F} dE  \int_{-\infty}^{\e_F}d\e_{\pp} &\int_{-\infty}^{\e_F} d\e_{\qp}  \dfrac{\delta(E - \varepsilon_{\pp} - \varepsilon_{\qp}+\varepsilon_p)}{(E-\e_l)^2} e^{-2B(E- \e_m)^2} + \\
& + \int_{-\infty}^{\e_F} d\e_m  \int_{\e_F}^{\infty} dE \int_{\e_F}^{\infty} d\e_{\pp} \int_{\e_F}^{\infty} d\e_{\qp}  \dfrac{\delta(E - \varepsilon_{\pp} - \varepsilon_{\qp}+\varepsilon_p)}{(E-\e_l)^2}e^{-2B(E- \e_m)^2}.
\end{align}
Shifting energies and going to new variables allows us to get rid of one integral by obtaining the complementary error function:
\begin{align}
I=\dfrac{2}{\sqrt{2B}\sqrt{\pi}}\int_{0}^{\infty}\int_{0}^{\infty} dE  d\e_{\pp} \left( \int_{-\infty}^{0} d\e_{\qp}  \dfrac{\delta(-E + \varepsilon_{\pp} - \varepsilon_{\qp}+\varepsilon_p)}{(E+\e_l)^2}+ \int_{0}^{\infty} d\e_{\qp}  \dfrac{\delta(E - \varepsilon_{\pp} - \varepsilon_{\qp}+\varepsilon_p)}{(E-\e_l)^2} \right)\text{erfc}(\sqrt{2B}E). 
\end{align}
In both cases the limits of the energy integrations can be restricted by an approximate evaluation of the delta function $\delta(E - \varepsilon_{\pp} - \varepsilon_{\qp}+\varepsilon_p)$. It gives the limit for the integration over $\e_{\qp}$ and $\e_{\pp}$ which is straightforward to calculate and plugging the result into the initial expression for $\delta S$ we obtain
\begin{align}\label{eq:step_back}
 \delta S &= - \dfrac{3g^2U^2 \rho}{\pi^2\sqrt{2\pi}} \int_{\lb\frac{1}{\Lambda}\rb^2}^{\lb\frac{l}{\Lambda}\rb^2} dB  \dfrac{1}{B\sqrt{B}}\int^{\infty}_{-\infty} d \tilde{\e_l} 
\int_{0}^{\infty} d\tilde{E}  \dfrac{\tilde{E}^2}{(\tilde{E}+\tilde{\e_l})^2}\text{erfc}(\tilde{E}).
\end{align}
\section{Divergence analysis of the Eq.~\ref{eq:entropy_correction}}
\label{appendix:div}

Let us take a closer look to the ``problematic" part of the Eq. \ref{eq:step_back} which would be the region of negative energies $\e$

\begin{align}\label{eq:integral}
I &= \int_{\lb\frac{1}{\Lambda}\rb^2}^{\lb\frac{l}{\Lambda}\rb^2} dB  \dfrac{1}{B\sqrt{B}}\int^{\infty}_{0} d\e \int_{0}^{\infty} d E  \dfrac{E^2}{(E-\e)^2}\text{erfc}(E) .
\end{align}

As was said in the main text, we want to act iteratively where each iteration presumes diagonalising the Hamiltonian up to some extend $B^*$. In order to do so, we need to look at the same expression in the case when the flow parameter of diagonalising the Hamiltonian is sent  not to  $\infty$, but only up to some value $B^*$. For this, we just need to remember the functional dependence of $M(B)$ and also the fact that $\delta S$ and hence also $I$ is proportional to $M^2$. We recall
\begin{align}
 M^l_{\pp\qp p}(B^*) &= U\rho\int^{\infty}_{-\infty} dE\dfrac{ \delta(E - \e_\pp - \e_\qp + \e_p)}{(E-\e_l)} \lb 1 - e^{-B^*(E-\e_l)^2} \rb
\end{align}
and see that the only difference between $M(B)$ and $M(\infty)$ is the pre-factor $ \lb 1 - e^{-B^*(E-\e_l)^2} \rb$ which we need to insert now to the $I$. By tracking all the changes we have done to the variables $E, \e_l$ during the calculation of the integrals we arrive at the following result for the final integral
\begin{align}
I&= \int_{\lb\frac{1}{\Lambda}\rb^2}^{\lb\frac{l}{\Lambda}\rb^2} dB  \dfrac{1}{B\sqrt{B}}\int^{\infty}_{0} d\e \int_{0}^{\infty} d E  \dfrac{E^2}{(E-\e)^2}\text{erfc}(E)  \lb 1 - e^{-\frac{B^*}{B}(E-\e)^2} \rb^2.
\end{align}

We can Taylor-expand the exponent which converges and cancel the problematic part
\begin{align}
I\approx \int_{\lb\frac{1}{\Lambda}\rb^2}^{\lb\frac{l}{\Lambda}\rb^2} dB  \dfrac{(B^*)^2}{B^3\sqrt{B}}\int^{\infty}_{0} d\e \int_{0}^{\infty} d E E^2\text{erfc}(E) (E-\e)^2
\end{align}
and integrate over $E$
\begin{align}
I\approx \int_{\lb\frac{1}{\Lambda}\rb^2}^{\lb\frac{l}{\Lambda}\rb^2} dB  \dfrac{(B^*)^2}{B^3\sqrt{B}}\int^{\infty}_{0} d\e\left(-\dfrac{3}{8}\varepsilon + \dfrac{6+5\varepsilon^2}{15\sqrt{\pi}}\right).
\end{align}
However, we need to integrate $\e$ only up to the bandwidth $\sqrt{2B}\e_0$. Here we do not specify $\e_0$ because we are not interested in the exact value of the integral -- only in its scaling with the subsystem size. By doing so we get
\begin{align}
I\approx \int_{\lb\frac{1}{\Lambda}\rb^2}^{\lb\frac{l}{\Lambda}\rb^2} dB  \dfrac{(B^*)^2}{B^3\sqrt{B}}\left(\dfrac{2\sqrt{2}}{5\sqrt{\pi}}\sqrt{B} - \dfrac{3\varepsilon_0^2}{8}B + \dfrac{2\sqrt{2}\varepsilon^3}{9\sqrt{\pi}}B^{3/2}\right),
\end{align}
where $B^*$ is some unknown function of $B$. However, one can see that the leading in $l$ contribution will come from the term

\begin{align}
I_{\text{lead}} \sim \int_{\lb\frac{1}{\Lambda}\rb^2}^{\lb\frac{l}{\Lambda}\rb^2} dB \left(\dfrac{B^*}{B}\right)^2.
\end{align}
We see that in case when $B^* \sim B^\alpha$ we obtain
\begin{align}
I_{\text{lead}} \sim \int_{\lb\frac{1}{\Lambda}\rb^2}^{\lb\frac{l}{\Lambda}\rb^2} dB B^{2(\alpha-1)} \sim \dfrac{\Lambda^{2-4\alpha}}{2\alpha-1}\left(\dfrac{1}{l^{2-4\alpha}} - 1\right).
\end{align}
In particular case of $\alpha = 0.25$ we have $1/l$ scaling hence making the correction term sub-leading.
\end{widetext}
\end{document}